# Physically Interpretable Feature Learning and Inverse Design of Supercritical Airfoils

Runze Li,*  Yufei Zhang,†  Haixin Chen‡

(*Tsinghua University, Beijing, 100084, People's Republic of China*)

**Machine-learning models have demonstrated a great ability to learn complex patterns and make predictions. In high-dimensional nonlinear problems of fluid dynamics, data representation often greatly affects the performance and interpretability of machine learning algorithms. With the increasing application of machine learning in fluid dynamics studies, the need for physically explainable models continues to grow. This paper proposes a feature learning algorithm based on variational autoencoders, which is able to assign physical features to some latent variables of the variational autoencoder. In addition, it is theoretically proved that the remaining latent variables are independent of the physical features. The proposed algorithm is trained to include shock wave features in its latent variables for the reconstruction of supercritical pressure distributions. The reconstruction accuracy and physical interpretability are also compared with those of other variational autoencoders. Then, the proposed algorithm is used for the inverse design of supercritical airfoils, which enables the generation of airfoil geometries based on physical features rather than the complete pressure distributions. It also demonstrates the ability to manipulate certain pressure distribution features of the airfoil without changing the others.**

---

* Ph. D., School of Aerospace Engineering, email: lirunze@tsinghua.edu.cn

†Associate professor, School of Aerospace Engineering, senior member AIAA, email: zhangyufei@tsinghua.edu.cn

‡ Professor, School of Aerospace Engineering, associate fellow AIAA, email: chenhaixin@tsinghua.edu.cn (Corresponding Author)

## Nomenclature

| | |
|---|---|
| $c$ | = physical code |
| $c_0$ | = sample label |
| $C_L$ | = lift coefficient |
| $C_p$ | = pressure coefficient |
| $H$ | = entropy |
| $p$ | = probability density |
| $M_w$ | = wall Mach number |
| $M_{w,1}$ | = wall Mach number in front of shock wave |
| $M_\infty$ | = free-stream Mach number |
| $\boldsymbol{\mu}$ | = mean vector |
| $Re$ | = Reynolds number |
| $\boldsymbol{\sigma}$ | = standard variance vector |
| $\boldsymbol{\Sigma}$ | = covariance matrix |
| $t_{max}$ | = maximum relative thickness |
| $\boldsymbol{v}$ | = data feature |
| $\boldsymbol{x}$ | = input vector |
| $X$ | = location |
| $X_1$ | = shock wave location |
| $\boldsymbol{z}$ | = latent variables |

## I. Introduction

Machine learning tools have been widely utilized in fluid dynamics studies and aerodynamic optimization designs. Since fluid dynamics problems are often high-dimensional and nonlinear, the performance of machine learning is heavily dependent on the choice of data representation (or features) [1]. There are two major types of data representations in machine learning applications for fluid dynamics problems. The first one is the traditional physical features, such as the combination and variation of fluid properties (e.g., mass flow, pressure gradient, shear rate tensor, etc.), or the features of flow structures (e.g., shock wave location, boundary layer thickness, vortex strength, etc.). The other type of representation is the data features that are directly extracted from data using machine learning algorithms, such as principal components analysis (PCA) and autoencoder (AE) [1].

Physical features have explicit mathematical expressions and clear physical meanings. Hence, they are usually utilized to help machine learning tools better capture the physical nature. For example, a complete tensor basis was constructed to represent the Reynolds stress in data-driven turbulence modeling, which contributed to an explainable and extrapolatable model [2]. Flow structure features were used in statistical studies to discover interpretable aerodynamic rules for airfoil optimization design [3]. Such features were also used as state values in reinforcement learning to capture general aerodynamic design policies that could be transfer-applied to different cases [4]. However, although flow structure features describe the most important information in the flow field, much information is still lost. Therefore, it is usually not enough to build machine learning models using only flow structure features.

Recently, data features have been widely used for dimensionality reduction in machine learning applications. PCA is a simple, linear and by far the most popular tool for dimensionality reduction [5]. It reduces the data frame by orthogonally transforming the data into a set of principal components. By choosing the top few principal components that explain most of the variation in the data, the dimensionality is reduced. AEs are a branch of neural networks that compress the input variables into a reduced dimensional space and then recreate the input dataset. In addition, convolutional autoencoders (CAEs) have been developed to extract features of two-dimensional flow field data with the help of convolutional layers [6].

However, many data features do not have clear physical meanings and this leads to black-box models. Although these models can accurately predict flow fields or improve turbulence models, researchers are increasingly unsatisfied with their non interpretability. There are various definitions of interpretability in machine learning, and one popular definition consists of making sense of the obtained model [7]. Generally, to interpret means "to explain the meaning of" or "present in understandable terms". In this paper, physical interpretability means that the features are presented in understandable physical terms, e.g., flow structure features.

Linear dimensionality reduction algorithms, such as PCA and dynamic mode decomposition (DMD) [8], decompose the entire flow field into different modes. Although the modes are interpretable, they usually cannot uniquely correspond to flow structure features, and their physical interpretability is limited. On the other hand, nonlinear algorithms have higher dimensionality reduction capabilities, but many of them (such as AE) also cannot extract physically interpretable features [1]. Compared with AE, generative models such as variational autoencoders (VAE) and generative adversarial networks (GAN) can obtain more interpretable features (i.e., latent variables for generative models). For example, beta-VAE [9,10] can learn disentangled features without supervision. Some complex generative models, such as StyleGAN [11,12] and deep convolutional inverse graphics networks (DCIGNs) [13], can learn disentangled and interpretable features by manipulating the training procedure. These generative models improve their interpretability by learning disentangled features because disentangled features are generally considered to contain interpretable information and reflect separate factors of variation in the data [10].

In summary, physically interpretable features in fluid dynamics studies should meet the following three requirements: physical explainability, completeness, and disentanglement, i.e.,

1. the features are explainable in fluid dynamics terms, e.g., flow structure features;

2. the features contain most of the information about the flow field, i.e., the flow field can be accurately reconstructed by a given combination of features;

3. the features are disentangled, i.e., different features describe different physical phenomena, and each physical phenomenon can be manipulated without affecting others.

Flow structure features do not meet the last two requirements because they contain only partial information, and sometimes there are also dependencies between features. Data features extracted by beta-VAE are learned without supervision and so they do not meet the first requirement. StyleGAN and DCIGN are designed to learn disentangled and interpretable features, but they have complex models and lack theoretical proof.

This paper proposes a novel physically interpretable feature learning algorithm based on VAE, denoted by PIVAE. The proposed algorithm divides its features (i.e., latent variables $z$ of VAE) into physical codes $c$ and data features $v$, i.e., $z = (c, v)$. The physical codes are assigned by users to capture specific flow structure features, and the data features will then automatically represent the rest of the information in the flow field. The physical codes and data features are trained to be independent of each other so that the proposed algorithm meets the above three requirements. Therefore, the proposed PIVAE can be used to reduce the dimensionality of the flow field while providing explicit features of flow structures, which can improve the interpretability of machine learning models built on the extracted features.

This paper is organized as follows. First, algorithms of VAE and the proposed algorithm are introduced, and it is theoretically proved that the proposed algorithm meets the three requirements of physically interpretable features. Next, the proposed algorithm is used to extract physically interpretable features of supercritical wall Mach number distributions. The performance of the proposed algorithm is also compared with other types of VAEs. Then, the proposed algorithm is used for the inverse design of supercritical airfoils. The resulting model is able to accurately generate airfoil geometries and wall Mach number distributions according to the given free stream conditions and wall Mach number distribution features.

## II. Theoretical methods

### 2.1 Variational autoencoders

Autoencoders have an encoder-decoder architecture. The encoder maps the data $x$ to latent variables $z$; then, the decoder reconstructs the data. The encoder and decoder are trained using backpropagation for accurate reconstruction of the input. The latent space of a single layered AE with a linear activation function strongly resembles the eigenspace achieved by PCA. Adding nonlinearity makes AE capable of representing data in lower dimensions, but the latent space becomes much more uninterpretable and nonregularized. To use the decoder for generative purposes, the latent space needs to be sufficiently regular so that we can take a point randomly from that latent space and decode it to obtain new content.

VAE addresses the issue of nonregularized latent space by forcing the latent distribution $p(z)$ to be a standard high-dimensional normal distribution $N(\mathbf{0}, I_{n_z})$. This provides the generative capability to the entire latent space ($n_z$ is the dimension of latent space). Fig. 1 shows the architectures of the AE and VAE.

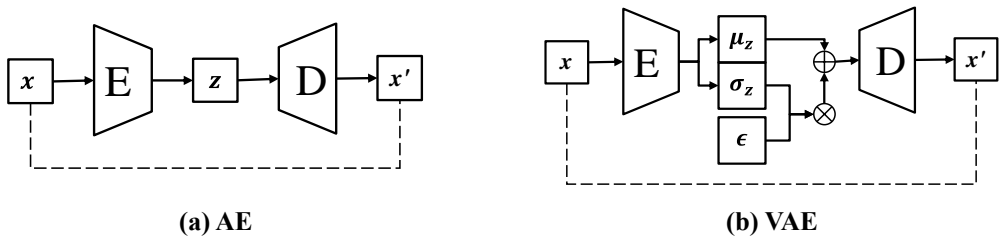

(a) AE  (b) VAE

**Fig. 1 Architectures of AE and VAE**

Consider some sample set $\{x_i\}$ $(i = 1, \cdots, N)$ that consists of N samples of independent and identically distributed (i.i.d.) random variables. The sample distribution is $p(x)$. Assume that the data $x$ are generated by a random process involving an unobserved continuous random variable $z$. Then, the process consists of two steps:

(1) a $z$ is sampled from a prior distribution $p(z)$; (2) a $x$ is sampled from the conditional distribution $p(x|z)$. $p(x|z)$ is called a probabilistic decoder, since given a code $z$, it produces a distribution over the possible corresponding values of $x$ [14]. Assume $p(x|z) = N(\mu_x, \sigma_x \odot I_{n_x})$, where $\mu_x$ is the mean vector, $\sigma_x$ is the standard variance vector, $I_{n_x}$ is the $n_x$-dimensional identity matrix, $n_x$ is the dimension of $x$, and $\odot$ is the Hadamard product (also known as the elementwise product). The covariance matrix of the normal distribution is a diagonal matrix because $x$ consists of i.i.d. random variables. Then, the loss function of the decoder minimizes the discrepancy between the input data and the reconstructed data, i.e.,

$$\max \ E_{z \sim q(z|x)} \log p(x|z) \tag{1}$$

The encoder of VAE is a recognition model $q(z|x)$ that approximates the intractable true posterior $p(z|x)$. It is also called a probabilistic encoder because given an $x$, it produces a distribution over $z$. Assume $q(z|x) = N(\mu_z, \Sigma_z)$, where $\mu_z$ is the mean vector and $\Sigma_z$ is the covariance matrix. With the commonly applied reparameterization trick [14], $z \sim \mu_z + \sigma_z \odot \epsilon$, where $\sigma_z$ is the standard variance vector and $\epsilon$ is the vector of sample noise ($\epsilon \sim N(0, I_{n_z})$). Then, the loss function of the encoder minimizes the difference between $q(z|x)$ and $p(z|x)$, which is described by the Kullback–Leibler divergence (KLD), i.e.,

$$\min \ \mathrm{KL}[q(z|x) \| p(z|x)] \tag{2}$$

Eq. (2) is intractable and is thus transformed into the computationally tractable evidence lower bound (ELBO) in VAE, which is described as Eq. (3).

$$\max \ L(x) = E_{z \sim q(z|x)} \log p(x|z) - \mathrm{KL}[q(z|x) \| p(z)] \tag{3}$$

The first term in Eq. (3) is the loss function of the decoder, and the second term encourages the approximate posterior to be close to the prior $p(z) = N(0, I_{n_z})$. Therefore, the loss function of VAE is the ELBO described in Eq. 3.

**2.2 Variational autoencoders with physical codes**

To learn physically interpretable features with VAE, the latent variables $z$ of the original VAE are divided into two parts, i.e., $z = (c, v)$. The latent variables $c$ are referred to as physical codes because they are used to capture specific physical features so that the first requirement of physically interpretable features is achieved. $v$ are called data features because they are trained to represent the rest of the information in $x$ so that the second requirement can be achieved. Then, the key issue is the third requirement, i.e., how to ensure that $(c, v)$ reconstructs $x$ accurately, while $c$ and $v$ are independent of each other.

Assume $p(v) = N(0, I_{n_v})$, as in VAE. Assume $p(c) = 1$ ($c \in [0,1]$), $c$ and $v$ are independent of each other. $n_v$ and $n_c$ are the dimensions of $v$ and $c$, respectively. The user-defined physical features extracted from $x$ are denoted by $c_0$, which become the ground truth labels of the sample. Denote $c_0 = f(x)$. In this study, the components of $c_0$ are selected from $\{M_\infty, C_L, X_1, M_{w,1}\}$, which are the free stream Mach number, airfoil lift coefficient, shock wave location, and wall Mach number in front of the shock wave. The detailed definitions will be introduced in the next section. Then, $p(c_0)$ is the label distribution of the sample set $\{x_i\}$.

First, the physical code $c$ is trained to represent the label $c_0$. Therefore, the loss function is

$$\max E_{c_0 \sim p(c_0|x), x \sim p(x)} \log q(c_0|x) \tag{4}$$

where $p(c_0|x)$ is the label distribution according to a given $x$. Since $c_0$ is explicitly extracted from $x$, $p(c_0|x)$ is a Dirichlet distribution. In other words, Eq. (4) becomes

$$\max E_{x \sim p(x)} \log q(f(x)|x) \tag{5}$$

Next, the ELBO becomes

$$\max L(x) = E_{v \sim q(v|x), c \sim q(c|x)} \log p(x|c, v) - \text{KL}[q(z|x) \| p(z)] \tag{6}$$

where the first term is the same as the loss function of the VAE encoder. The second term becomes

$$\text{KL}[q(z|x) \| p(z)] = \text{KL}[q(v|x)q(c|x) \| p(v)p(c)] \tag{7}$$

where $q(z|x) = q(v|x)q(c|x)$ holds because of the underlying assumption of the reparameterization trick. Then, Eq. (7) becomes

$$\text{KL}[q(z|x) \| p(z)] = \mathop{E}_{\substack{v \sim q(v|x) \\ c \sim q(c|x)}} \log[q(v|x)q(c|x)] - \mathop{E}_{\substack{v \sim q(v|x) \\ c \sim q(c|x)}} \log[p(v)p(c)] \tag{8}$$

$$= \mathop{E}_{v \sim q(v|x)} \log[q(v|x)] + \mathop{E}_{c \sim q(c|x)} \log[q(c|x)] - \mathop{E}_{v \sim q(v|x)} \log p(v) \tag{9}$$

$$= \text{KL}[q(v|x) \| p(v)] + H(q(c|x)) \tag{10}$$

With the reparameterization trick, $q(\boldsymbol{v}|\boldsymbol{x})$ is a normal distribution described by $\boldsymbol{v} \sim \boldsymbol{\mu}_v + \boldsymbol{\sigma}_v \odot \boldsymbol{\epsilon}$ and $q(\boldsymbol{c}|\boldsymbol{x})$ is a normal distribution described by $\boldsymbol{c} \sim \boldsymbol{\mu}_c + \boldsymbol{\sigma}_c \odot \boldsymbol{\epsilon}$, $\boldsymbol{\epsilon} \sim N(\boldsymbol{0}, \boldsymbol{I})$. Then, the KLD loss is

$$\mathrm{KL}[q(\boldsymbol{v}|\boldsymbol{x}) \| p(\boldsymbol{v})] = \frac{1}{2}\sum_{j=1}^{n_v}\left[1 + \log \sigma_{v,(j)}^2 - \mu_{v,(j)}^2 - \sigma_{v,(j)}^2\right] \tag{11}$$

, where $\mu_{v,(j)}$ is the j$^{\text{th}}$ component of $\boldsymbol{\mu}_v$ and $\sigma_{v,(j)}$ is the j$^{\text{th}}$ component of $\boldsymbol{\sigma}_v$. The entropy of $q(\boldsymbol{c}|\boldsymbol{x})$ becomes

$$H\big(q(\boldsymbol{c}|\boldsymbol{x})\big) = \frac{n_v}{2}(\log 2\pi + 1) + \frac{1}{2}\sum_{j=1}^{n_v} \log \sigma_{c,(j)}^2 \tag{12}$$

which tries to minimize the variance of $q(\boldsymbol{c}|\boldsymbol{x})$. In other words, $q(\boldsymbol{c}|\boldsymbol{x})$ is encouraged to be as close as possible to the Dirichlet distribution $p(\boldsymbol{c}_0|\boldsymbol{x})$. In practice, the proposed algorithm neglects $\boldsymbol{\sigma}_c$ and directly assumes a Dirichlet distribution $q(\boldsymbol{c}|\boldsymbol{x})$. Then, the loss function of the physical codes (Eq. (5)) becomes

$$\min \sum_{i=1}^{N} \|\boldsymbol{c}_i - \boldsymbol{c}_{0,i}\| \tag{13}$$

where $\boldsymbol{c}_i$ is the physical codes of sample $\boldsymbol{x}_i$ predicted by the encoder and $\boldsymbol{c}_{0,i} = \boldsymbol{f}(\boldsymbol{x}_i)$ is the ground truth label.

Next, the architecture of the variational autoencoder with physical codes (PIVAE) is shown in Fig. 2. The loss function is

$$\min \sum_{i=1}^{N} \|\boldsymbol{c}_i - \boldsymbol{c}_{0,i}\| - L(\boldsymbol{x}) \tag{14}$$

$$L(\boldsymbol{x}) = E_{v \sim q(v|x)} \log p(\boldsymbol{x}|\boldsymbol{c}, \boldsymbol{v}) - \mathrm{KL}[q(\boldsymbol{v}|\boldsymbol{x}) \| p(\boldsymbol{v})] \tag{15}$$

It should be noted that PIVAE assumes $p(\boldsymbol{c}) = 1$ while encouraging $\boldsymbol{c} = \boldsymbol{c}_0$. Therefore, the sample set needs to be manipulated to have its label distribution $p(\boldsymbol{c}_0) = 1$, so that Eq. 9 holds, and $\boldsymbol{c}$ and $\boldsymbol{v}$ can become independent of each other.

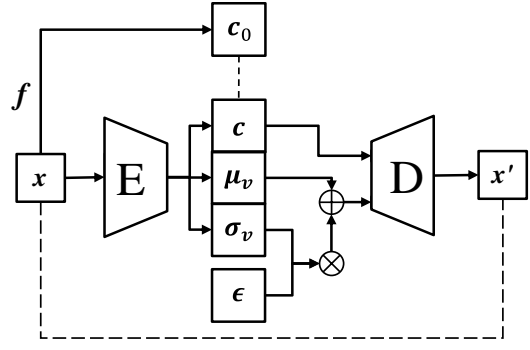

**Fig. 2 Architecture of PIVAE**

The PIVAE algorithm is shown in Algorithm A. To improve the reconstruction ability of PIVAE, an additional term $\text{loss}_{\text{mean}}$ is added to the loss function in Eq. (14). It is the loss function of the traditional autoencoder that encourages the model to accurately reconstruct the normalized data $x$ with the predicted physical code $c$ and the mean data feature $\mu_v$. The loss function $\text{loss}_{\text{re}}$ is the estimation of the first term in Eq. (15), which has the same form as Eq. (1) of VAE [14]. The 're' stands for reparameterization. The ratio $r$ in the 12th line slowly increases from a small number to a constant value so that the algorithm is converges better. The model is updated by the Adam optimizer [20].

Algorithm A: PIVAE

| | |
|---|---|
| 1 | Gather samples $\{x_i\}$ $(i = 1, \ldots, N)$ and the corresponding labels $\{c_{0,i}\}$. |
| 2 | for epoch = 1, …, $N_{\text{epoch}}$; { |
| 3 | for k = 1, …, $N_{\text{mini-batch}}$; { |
| 4 | $c_i, \mu_{v,i}, \sigma_{v,i} = \text{Encoder}(x_i)$ $(i = 1, \ldots, N_k)$; ($N_k$ samples in kth mini-batch;) |
| 5 | $\text{loss}_{\text{code}} = \sum_{i=1}^{N_k} \|c_i - c_{0,i}\|/N_k$; |
| 6 | $\bar{x}_i = \text{Decoder}(c_i, \mu_{v,i})$; (Mean data reconstructed by $c_i$ and $\mu_{v,i}$;) |
| 7 | $\text{loss}_{\text{mean}} = \sum_{i=1}^{N_k} \|\bar{x}_i - x_i\|/N_k$; |
| 8 | $v_i \sim \mu_{v,i} + \sigma_{v,i} \odot \epsilon$, $\epsilon \sim N(0, I)$; (Reparameterization;) |
| 9 | $\tilde{x}_i = \text{Decoder}(c_i, v_i)$; |
| 10 | $\text{loss}_{\text{re}} = \sum_{i=1}^{N_k} \|\tilde{x}_i - x_i\|/N_k$; ('re' stands for reparameterization;) |
| 11 | $\text{loss}_{\text{kld}} = \sum_{i=1}^{N_k} \text{KL}[q(v_i|x_i)\|p(v_i)]/N_k$ (refers to Eq. (11)); |
| 12 | $\text{loss} = \text{loss}_{\text{code}} + \text{loss}_{\text{mean}} + r(0.1\,\text{loss}_{\text{re}} + \text{loss}_{\text{kld}})$; |
| 13 | Update model with the Adam optimizer; |
| 10 | } end; |

## 2.3 Comparison with conditional variational autoencoders

To learn physically interpretable features with VAE, one's first thought is to use conditional variational autoencoders (cVAE). cVAE can use user-defined physical features $c_0$ as part of the input of the encoder and decoder. The architecture of cVAE is shown in Fig. 3(a). The loss function is

$$\max L(x) = E_{v \sim q(v|x,c_0)} \log p(x|c_0, v) - \text{KL}[q(v|x,c_0) \| p(v)] \quad (16)$$

It can be seen that the loss function of cVAE (Eq. (16)) greatly resembles Eq. (15). Furthermore, if the conditional recognition model $q(z|x, c_0)$ is replaced by $q(z|x)$ (Fig. 3(b)), which is the approximation made in Gaussian stochastic neural networks (GSNN) [15], Eq. (16) becomes

$$\max L(x) = E_{v \sim q(v|x)} \log p(x|c_0, v) - \text{KL}[q(v|x) \| p(v)] \quad (17)$$

Then, the loss function of Eq. (17) is the same as that of a well-trained VAS-code model, in which the physical codes $c$ can be considered to be equal to the label $c_0$. The architecture of GSNN also resembles that of PIVAE. Therefore, cVAE (or GSNN) can generate similar physically interpretable features as PIVAE, as long as its training sample set has a uniform label distribution.

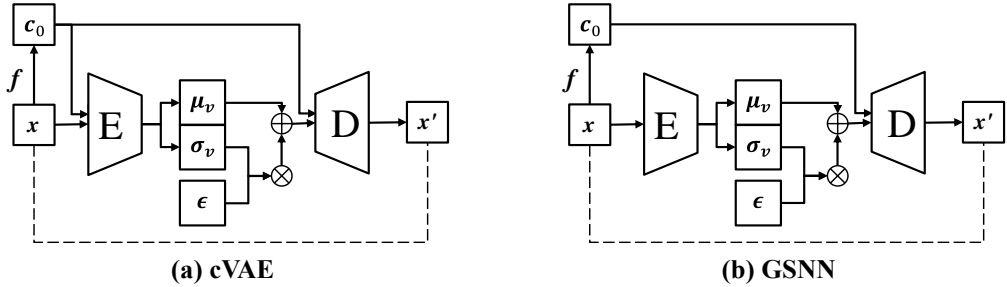

(a) cVAE  (b) GSNN
Fig. 3 Architectures of cVAE and GSNN

### III. Feature learning of supercritical airfoils

#### 3.1 Geometry, simulation and physical features of supercritical airfoils

The supercritical airfoils are constructed by the class shape transformation (CST) method [16] with a ninth-order Bernstein polynomial as the shape function. The airfoils have a unit chord length. A C-grid is used for computational fluid dynamics (CFD) simulation with an open-source solver, CFL3D [17]. $\Delta y+$ of the first grid layer is always set to be less than one. 301 grid points are distributed on the airfoil surface. The MUSCL scheme, Roe's scheme, the lower-upper symmetric Gauss-Seidel method, and the $k - \omega$ shear stress transport (SST) model are used in Reynolds average Navier–Stokes (RANS) simulations. The CFL number is 2.0 for 8000 steps.

Fig. 4 a) shows the experimental pressure coefficient ($C_p$) distributions of an RAE2822 airfoil [18] compared with the CFD results of different grid sizes. The free-stream Mach number ($M_\infty$) is 0.725, the Reynolds number ($Re$) based on the unit chord length is $6.5 \times 10^6$, and the angle of attack ($AoA$) is 2.55 degrees. The three grids, with sizes of 20,000, 40,000, and 80,000, respectively. All grids can achieve a similar $C_p$ resolution. The medium grid is shown in Fig. 4 b), which is employed in this paper.

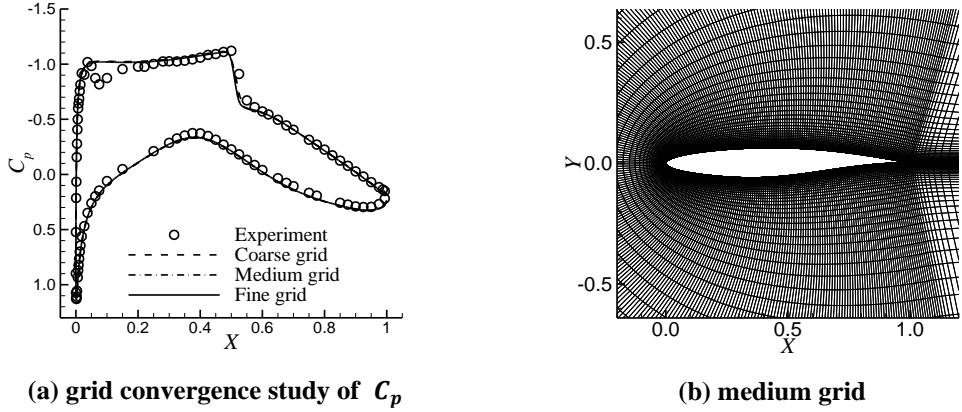

(a) grid convergence study of $C_p$      (b) medium grid

**Fig. 4 CFD validation and grid**

In this section, wall Mach number distributions of single shock wave supercritical airfoils construct the sample set $\{x_i\}$. The wall Mach number ($M_w$) is the Mach number calculated based on an isentropic relationship with the pressure coefficient ($C_p$) on the airfoil surface and the free-stream Mach number [19]. The location and strength of shock waves are selected for physically interpretable feature learning, and the features are shown in Fig. 5. The shock wave appears as an abrupt $M_w$ decrease on the upper surface, and the $M_w$ in front of the shock wave ($M_{w,1}$) exceeds one. The shock wave is first roughly located by the largest $-dM_w/dX$ location, $\tilde{X}$. Then, $M_{w,1}$ is defined as $M_w$ at location $X_1$ in front of $\tilde{X}$, where $dM_w/dX|_{X_1} = -1$ or $d^2M_w/dX^2$ reaches a local minimum.

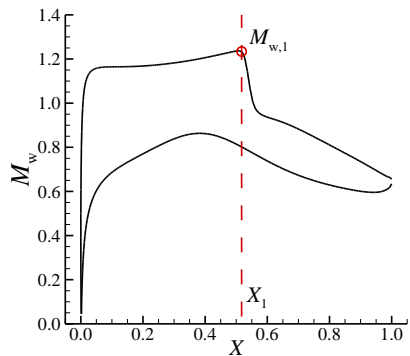

**Fig. 5 Definition of shock wave features**

### 3.2 Sample preparation

There are two sample sets used in this study. The first one is generated by perturbation of CST parameters to three baseline airfoils shown in Fig. 6, i.e., OAT15A, RAE2822, and another supercritical airfoil for dual-aisle civil aircraft (denoted by "dual-aisle" in Fig. 6). The CST parameters are perturbed by adding a random vector sampled from a normal distribution $N(0, \sigma I)$, where $\sigma$ is the standard variance that equals half the maximum absolute value of the baseline CST parameters. Each baseline airfoil is repeatedly perturbed and evaluated by computational fluid dynamics (CFD) under 18 different conditions. The 18 conditions are defined by all combinations of $M_\infty = \{0.71, 0.73, 0.75\}$, $C_L = \{0.60, 0.75, 0.90\}$, and $t_{max} = \{0.10, 0.12\}$. The Reynolds number based on the unit chord length equals $1 \times 10^7$. $t_{max}$ is the airfoil maximum thickness, and the airfoil geometry is scaled to match the specified $t_{max}$ during sampling. Only the airfoils that have a converged CFD result and a single shock wave are selected into the sample set along with the corresponding labels $\{M_\infty, C_L, M_{w,1}, X_1\}$. Over 200,000 CFD simulations are conducted, and 11,000 samples are eventually selected.

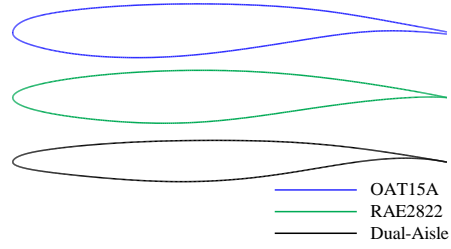

**Fig. 6 Baseline airfoils**

The second sample set is generated by an output space sampling algorithm (OSS) [3]. OSS is employed to generate a sample set of which the label distribution $p(c_0)$ approximates the prior distribution $p(c) = 1$. The second sample set contains 11,000 single shock wave airfoil samples, with the labels $c_0 = \{M_\infty, C_L, M_{w,1}, X_1\}$. The ranges of labels and $t_{max}$ are listed in Table 1. Fig. 7 shows the sample distribution in the label space of $\{M_\infty, M_{w,1}, X_1\}$. Each colored sphere represents the number of samples in its neighborhood in the label space. There are 45 spheres in Fig. 7 a) representing the label distribution of the first sample set generated by CST perturbation. The average number of samples for each sphere is approximately 250 if $p(c_0) = 1$. Fig. 7 b) has 60 spheres representing the label distribution of the second sample set generated by OSS. The average number of samples for each sphere is approximately 180 if $p(c_0) = 1$. Fig. 7 indicates that the sample generated by perturbation is concentrated in a small area, but the label distribution of the samples generated by OSS is much closer to a uniform distribution.

Table 1 Ranges of labels and $t_{max}$

| | $M_\infty$ | $C_L$ | $X_1$ | $M_{w,1}$ | $t_{max}$ |
|---|---|---|---|---|---|
| Range | [0.71,0.76] | [0.60,0.90] | [0.20,0.75] | [1.10,1.26] | [0.09,0.13] |

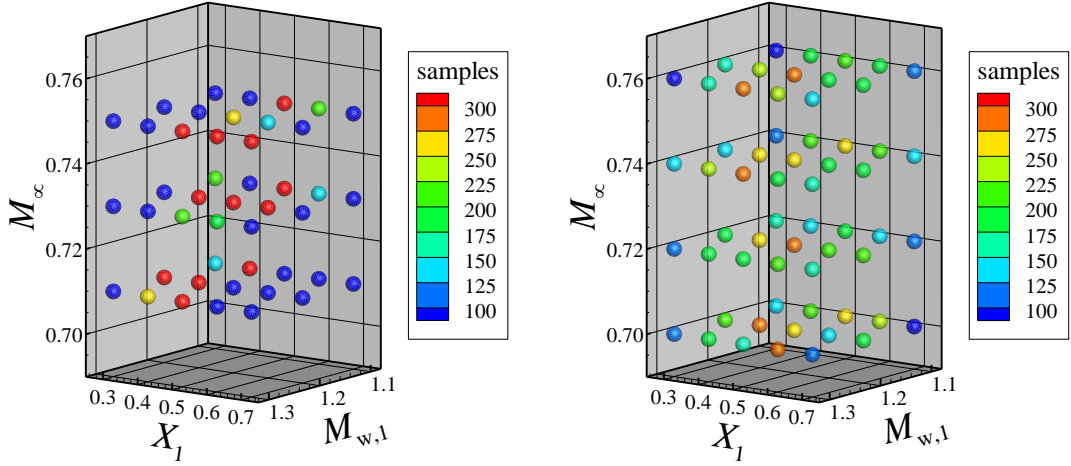

a) Samples generated by perturbation    b) Samples generated by OSS

Fig. 7 Sample distribution in the label space

### 3.3 Comparison between VAEs

In this section, the capability of generating physically interpretable features is compared between VAE, cVAE, and PIVAE using the second sample set. Then, the performance of PIVAE using the two sample sets is also compared. The physical code of PIVAE and the label of cVAE are selected to be $\{C_L, X_1\}$. The dimension of the data features of PIVAE is six, the dimension of latent variables of cVAE is six, and the dimension of latent variables of VAE is eight. The input data $x$ is the airfoil wall Mach number distribution that is redistributed from a fixed chordwise distribution, and the dimension of $x$ is $1 \times 401$. Fig. 8 shows the detailed architecture of the networks in PIVAE. The $k$ in the input and output is one in this section because the data contain only the wall Mach number distribution. $k$ will be two in section IV when the data contain the wall Mach number distribution and the airfoil y coordinates. The encoder and decoder are both fully connected networks, and the number of layers and the size of each layer are shown in Fig. 8. VAE and cVAE have a similar architecture to PIVAE; their encoder and decoder have the same number of layers and size of each layer, and the only difference is the dimensions of $c$ and $v$.

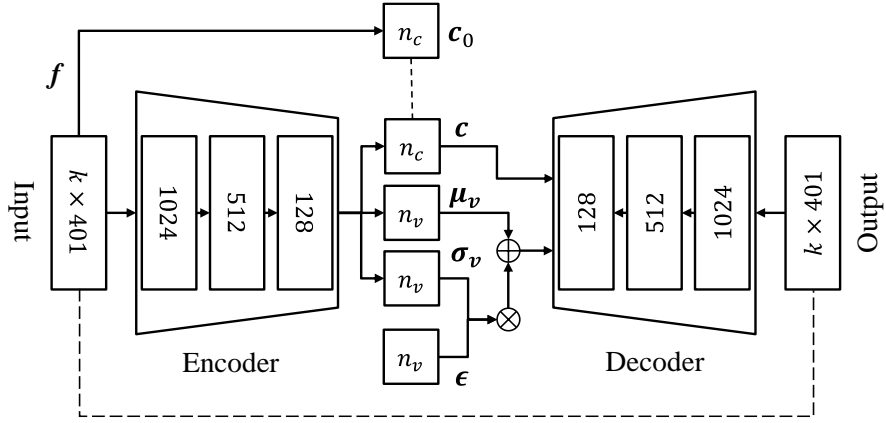

**Fig. 8 Detailed architecture of PIVAE networks**

A total of 1,000 samples are randomly selected for the test set, and the remaining 10,000 samples form the training set. The learning rate of the Adam optimizer is 0.0002 for 10,000 epochs, and the mini-batch size is 64. The weight of the KLD loss function gradually increases from 0.001 to 0.1 in 5,000 epochs, and the learning rate is reduced by a factor of 0.1 every 1,000 steps. Fig. 9 shows the loss function history of different models. The solid line without scatters in Fig. 9 a) is the reconstruction loss, and the solid line with scatters is the reconstruction loss on the test set. The reconstruction loss of PIVAE is $\text{loss}_{\text{mean}}$ in Algorithm A, and it is $\text{loss}_{\text{re}}$ for VAE and cVAE because they do not have the $\text{loss}_{\text{mean}}$ term. Fig. 9 b) shows the history of KLD loss. The $\text{loss}_{\text{code}}$ of PIVAE is eventually $O(10^{-5})$ after training.

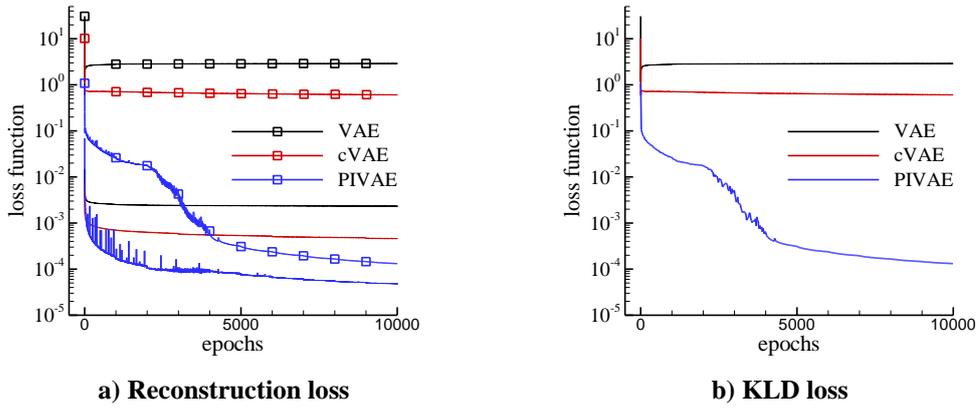

a) Reconstruction loss    b) KLD loss

**Fig. 9 History of the loss function**

Fig. 10 shows the reconstruction of wall Mach number distributions of samples from the training and test sets. Table 2 shows the ground truth labels of the selected samples in Fig. 10. The reconstruction is conducted by encoding the input data and decoding the mean value of latent variables, i.e., by skipping the reparameterization in VAEs. Four samples are randomly selected from the training and test sets. The first row in Fig. 10 is the samples from the training set, and the second row is the test set. The black solid line is the true value of the sample data, the green dashed line with square scatters is the reconstruction result of VAE, the red dashed line with triangle scatters is that of cVAE, and the blue dashed line is that of PIVAE. Fig. 10 shows that VAE has relatively low reconstruction accuracy, which can also be seen from Fig. 9(a). PIVAE has satisfactory reconstruction accuracy because it minimizes $\text{loss}_\text{mean}$, while VAE and cVAE minimize $\text{loss}_\text{re}$.

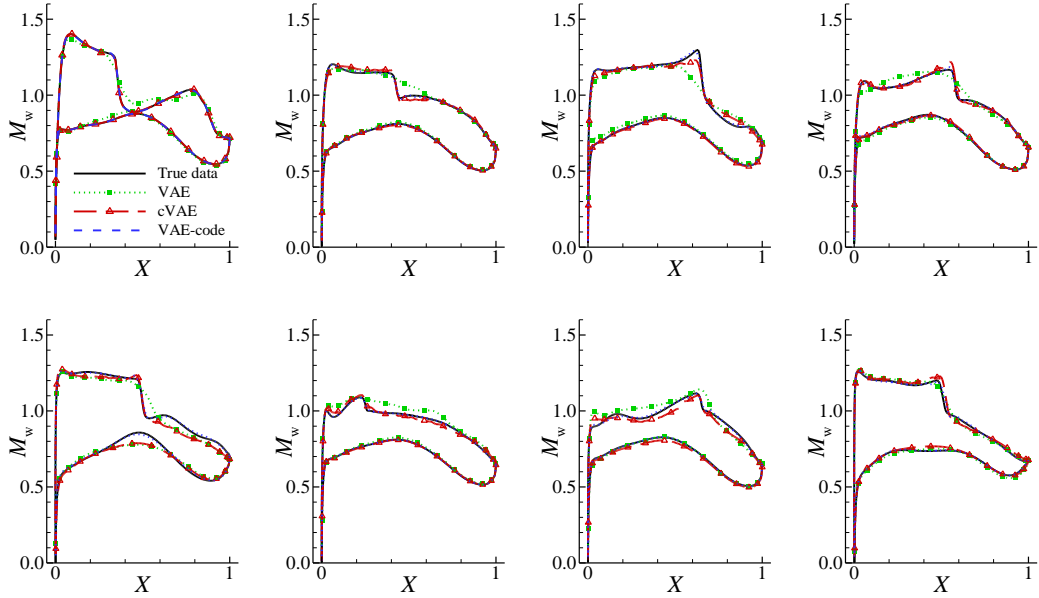

**Fig. 10 Reconstruction of samples in the training and test sets**
**(Black: true data; green: VAE; red: cVAE; blue: PIVAE)**

Table 2 Labels of the selected samples in Fig. 10

|  | $M_\infty$ | $C_L$ | $X_1$ | $M_{w,1}$ | $t_{max}$ |
|---|---|---|---|---|---|
| Training set 1 | 0.740 | 0.68 | 0.34 | 1.25 | 0.12 |
| Training set 2 | 0.715 | 0.80 | 0.41 | 1.13 | 0.12 |
| Training set 3 | 0.750 | 0.79 | 0.64 | 1.29 | 0.11 |
| Training set 4 | 0.725 | 0.66 | 0.56 | 1.16 | 0.13 |
| Test set 1 | 0.750 | 0.83 | 0.48 | 1.20 | 0.09 |
| Test set 2 | 0.710 | 0.60 | 0.24 | 1.08 | 0.11 |
| Test set 3 | 0.710 | 0.63 | 0.64 | 1.11 | 0.12 |
| Test set 4 | 0.730 | 0.81 | 0.51 | 1.08 | 0.09 |

There are three aspects in the physical interpretability of VAEs: 1) the reconstruction accuracy 2) the generation accuracy and 3) the perturbation disentanglement. The reconstruction accuracy describes the error of the physical codes extracted by the encoder, i.e.,

$$E_{x\sim p(x)}\|f(x) - c_{\text{encoder}}(x)\| \tag{18}$$

where $c_{\text{encoder}}$ denotes the physical codes $c$ of data $x$ predicted by the encoder and $x$ is the ground truth data. This is the same as the loss function of physical codes in PIVAE, i.e., $\text{loss}_{\text{code}}$ in Algorithm A. The accuracy is validated by the loss function history in Fig. 9(a). On the other hand, cVAE is not relevant to the reconstruction accuracy evaluation because it does not extract the physical codes by its encoder. The generation accuracy describes the error of decoder, i.e.,

$$E_{c\sim p(c), v\sim p(v)}\|c - f(x')\| \tag{19}$$

where $x'$ is generated by the decoder with the sampled $(c, v)$. It describes whether the generated wall Mach number distribution $x'$ achieves the physical features specified by physical codes $c$.

An intuitive demonstration of the generation accuracy is the airfoil matrix in Fig. 11. The wall Mach number distributions are generated by the decoder. Each component of the scaled physical codes $c = [C_L, X_1]$ varies from zero to one. The data features $v = 0$ indicate that the generated data are most likely to occur because the prior distribution $p(v)$ is a standard normal distribution. The red dashed lines show the shock wave location $X_1$ specified by the physical codes. Fig. 11 intuitively proves the generation accuracy of $X_1$ for the PIVAE. The generation accuracy of $C_L$ is difficult to discuss in Fig. 11 because the pressure coefficient distribution cannot be obtained without the free stream Mach number. Nevertheless, Fig. 11 shows that PIVAE generates wall Mach number distributions that achieve the physical features specified in the physical codes $c$.

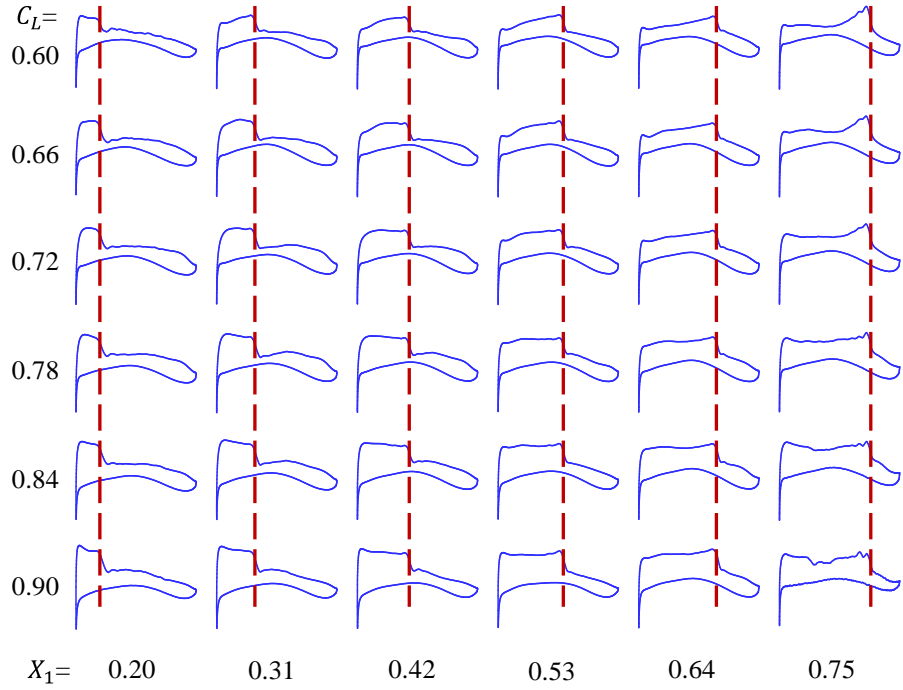

Fig. 11 Generation accuracy of physical code $X_1$ (PIVAE)

The perturbation disentanglement describes the disentanglement between the physical labels $c_0$ and latent variables $v$ in cVAE or the physical codes $c$ and the data features $v$ in PIVAE. Fig. 12 shows the generated wall Mach number distributions of PIVAE and cVAE. The black lines are the true wall Mach number distributions of the four selected samples from the test set, which are the same samples in the second row of Fig. 10. The blue lines are the reconstructed data. Each component of the scaled $c_0$ or $c$ is perturbed by 0.1 while keeping the other component and $v$ unchanged. The newly generated data are plotted as orange dashed lines in Fig. 12. These generated data probably do not appear in the training set. It shows that the physical features $C_L$ and $X_1$ are changed corresponding to the perturbation of physical codes, while the other parts of the wall Mach number distribution are mostly undisturbed.

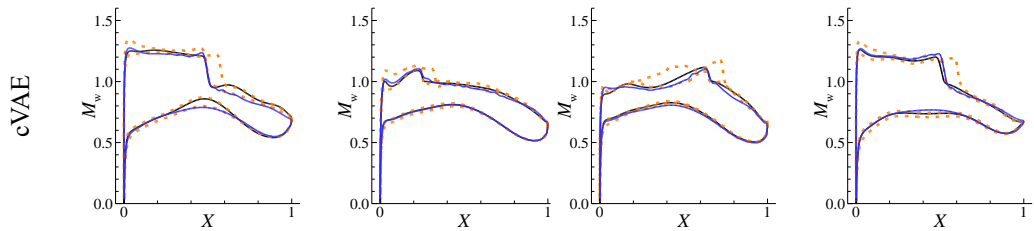

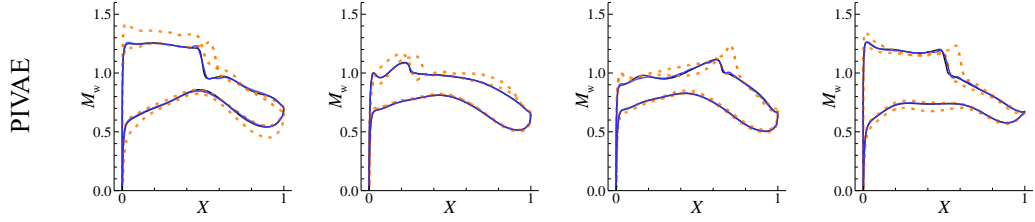

**Fig. 12 Perturbation of $c_0$ in cVAE and $c$ in PIVAE**
**(Black: true data; blue: reconstructed data; orange: generated data by perturbation of $c_0/c$)**

On the other hand, each component of $v$ in VAE, cVAE and PIVAE is perturbed while keeping the others unchanged to show the influence of each feature. The black solid lines are the true wall Mach number distributions, which are the same samples in the second row of Fig. 10. The blue solid lines are the reconstructed data. Only the first two samples in Fig. 12 are used because the dimension of $v$ is larger, and it is clearer to plot a few generated data in one subfigure. The first row of Fig. 13 shows the generated wall Mach number distributions (orange dashed line) of VAE, the second row is cVAE, and the third row is PIVAE. For VAE, it can be seen that three of the eight latent variables have a significant effect on the generated data because there are three distinguishable orange dashed lines in each figure of the first row. Although the generated data appear to be reasonable, the latent variables do not have clear and disentangled physical meanings. The last two rows of Fig. 13 show that all six components of $v$ of cVAE and PIVAE are effective. Fig. 13 indicates that the $v$ of cVAE and PIVAE is independent of the labels $c_0$ of cVAE and the physical codes $c$ of PIVAE, i.e., $\{C_L, X_1\}$.

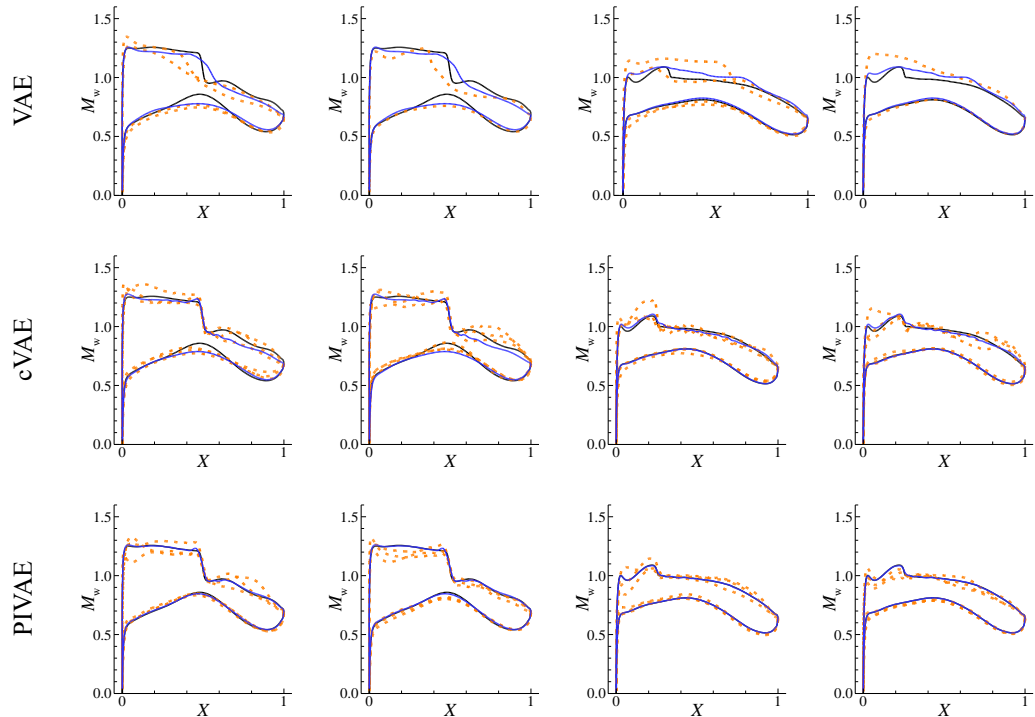

**Fig. 13 Perturbation of latent variables (VAE, cVAE) and data features (PIVAE)**

To further study the disentanglement between $c$ and $v$, 100 samples are randomly selected from the training set and test set. The 200 samples are reconstructed by VAEs. Each component of $v$ is perturbed by the same scale as the perturbation in Fig. 13. Wall Mach number distributions are decoded based on the perturbed $v$ (together with the fixed $c$ for cVAE and PIVAE). Then, the $\{C_L, X_1\}$ of 4,000 (= 200 × (8 + 6 + 6)) generated $M_w$ distributions are extracted and compared with the true value of the unperturbed sample. The change in $\{C_L, X_1\}$ for each generated sample relative to the unperturbed sample is plotted in Fig. 14. This proves that PIVAE has the best disentanglement in extracted features while the latent variables of VAE are entangled with $\{C_L, X_1\}$.

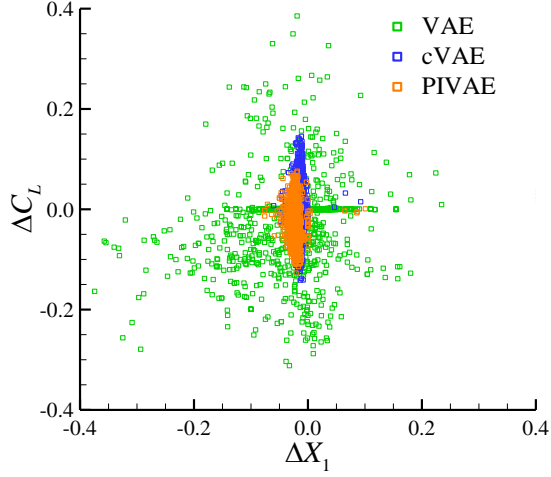

**Fig. 14 The change in $\{C_L, X_1\}$ during the perturbation of $v$**

Finally, the influence of the training set is discussed by comparing two PIVAE models trained on the two sample sets in Section 3.2. The results of the PIVAE model trained on the second sample set, i.e., the sample set with a uniform label distribution, are shown in Fig. 12 and Fig. 13. The results of the PIVAE model trained on the first randomly generated sample set are shown in Fig. 15. The black lines in the first row of Fig. 15 show four samples randomly selected from the test set. Their ground truth labels are listed in Table 3. The blue lines are the reconstruction result, and the orange lines are the newly generated data by the perturbation of physical codes. The second row shows the data generated by the perturbation of data features, and the results of the second and third samples are plotted. Fig. 15 shows that the physical codes $c$ correspond badly to $\{C_L, X_1\}$, and the generated wall Mach number distributions are also unreasonable.

In summary, physically interpretable features can be learned with PIVAE training on a sample set with uniform label distributions. cVAE can obtain similar results to PIVAE when being trained on the same sample set.

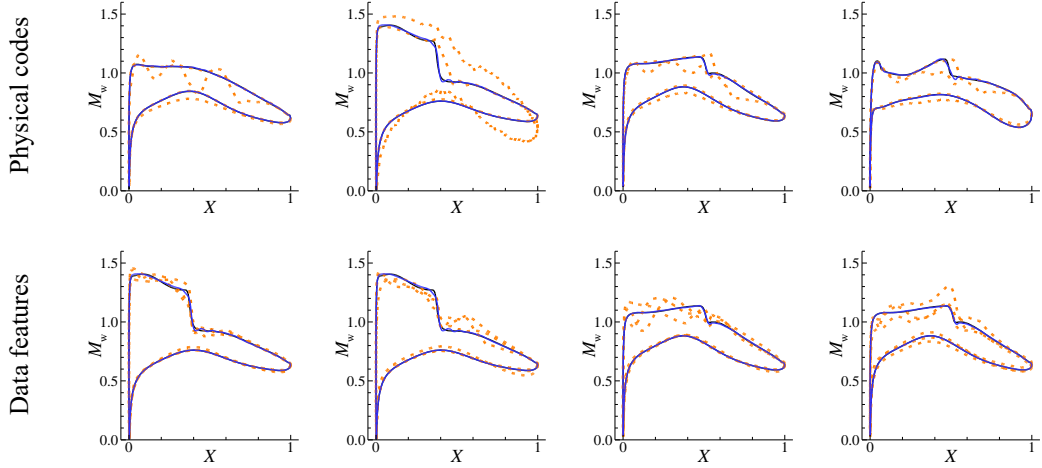

**Fig. 15 Perturbation of physical codes and data features (PIVAE trained on the first sample set)**

**(Black: true data; blue: reconstructed data; orange: generated data by perturbation of $c$ and $v$)**

Table 3 Labels of the selected samples in Fig. 15

|   | $M_\infty$ | $C_L$ | $X_1$ | $M_{w,1}$ | $t_{max}$ |
|---|---|---|---|---|---|
| 1 | 0.710 | 0.60 | 0.00 | 1.00 | 0.12 |
| 2 | 0.710 | 0.90 | 0.36 | 1.27 | 0.10 |
| 3 | 0.730 | 0.60 | 0.50 | 1.12 | 0.12 |
| 4 | 0.730 | 0.60 | 0.48 | 1.11 | 0.10 |

## IV. Inverse design of supercritical airfoils

One of the major challenges of inverse design is that it is difficult to produce reasonable wall Mach number distribution targets that have the desired physical features. For example, when a single shock wave airfoil with a specified $\{M_\infty, C_L, X_1, M_{w,1}\}$ is desired, it is difficult to manually draw a realizable and complete wall Mach number distribution for inverse design. To provide reasonable targets, inverse design methods based on VAEs have been developed [21]. VAEs were used to build reasonable wall Mach number distribution targets, and additional networks were built to predict the corresponding geometry, i.e., airfoil y coordinates. Unfortunately, the previous method [21] was still unable to specify the physical features of the generated wall Mach number distributions.

With the proposed PIVAE, reasonable wall Mach number distributions can be generated with specified physical features, and the corresponding airfoil geometry can be accurately reconstructed at the same time. This enables inverse design without building additional networks. The architecture of PIVAE for training inverse design models is shown in Fig. 16. The input data $x$ contain airfoil y coordinates as well as the $M_w$ values on the airfoil surface. The physical codes are specified to $\{M_\infty, C_L, X_1, M_{w,1}\}$ in this section. The training sample set and settings are the same as those of PIVAE in section III.

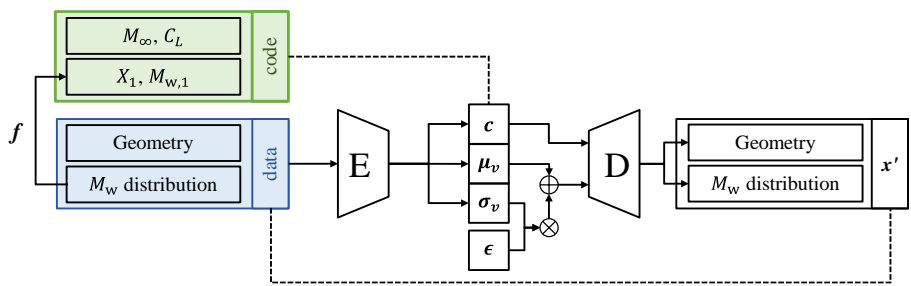

**Fig. 16 Architecture of PIVAE for training inverse design models**

After the PIVAE is trained, the decoder of the PIVAE is used for inverse design. Fig. 17 shows the architecture of PIVAE for inverse design applications. When airfoils with specified physical codes are desired, the physical codes and arbitrary data features are fed into the decoder. Then, the airfoil wall Mach number distribution and geometry are generated. If multiple airfoils are needed under the same physical codes, data features can be randomly valued without affecting the desired physical features.

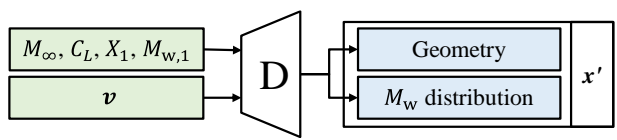

**Fig. 17 Architect of PIVAE for inverse design applications**

Fig. 18 and Fig. 19 demonstrate several matrices of generated wall Mach number distributions and geometries under different physical codes $c$. The data features $v$ are zero. Only two components of the physical codes are changed in each matrix, and the unscaled values are plotted on either side of the matrix. The dashed lines are the data generated by PIVAE. These data are not necessarily in the training set. The generated geometries (black solid lines in Fig. 18 b)) are evaluated for CFD validation, and the wall Mach number distributions are extracted from the CFD results and plotted as colored solid lines. The blue line indicates that the generated data match well with the CFD validation, the green line indicates minor discrepancy, and the red line indicates large difference or undesired results. This shows that PIVAE has poor inverse design ability at matrix corners, which is inevitable to some extent because these corners are probably physically unrealizable. Other than these corners, PIVAE shows very good ability to generate airfoils with any specified physical features. If multiple airfoils are desired for the same physical codes, the data features $v$ can be randomly sampled to generate different airfoils.

Fig. 18 shows the airfoil matrix of varying $M_{w,1}$ and $X_1$; the other two physical codes are $M_\infty = 0.73$ and $C_L = 0.75$. This shows that the PIVAE can generate single shock wave supercritical airfoils at a specified free stream condition, while the shock wave features can be manipulated. Fig. 19 shows the airfoil matrix when the free stream condition is changed. This shows that the PIVAE is effective in a wide range of free stream conditions, which covers most of the flight conditions of transonic civil aircraft.

On the other hand, Fig. 12 and Fig. 18 also indicate that the PIVAE has the ability to manipulate the physical features of a given airfoil. When the geometry and wall Mach number distribution of an airfoil are provided, the physical codes and data features can be extracted by the encoder. Then, if a certain physical feature is to be manipulated, new airfoils can be generated with the modified physical codes and the original data features. This is the same procedure as generating the airfoil matrices in Fig. 18. Then, the other features of airfoil geometry and wall Mach number distribution can remain unchanged to the greatest possible extent.

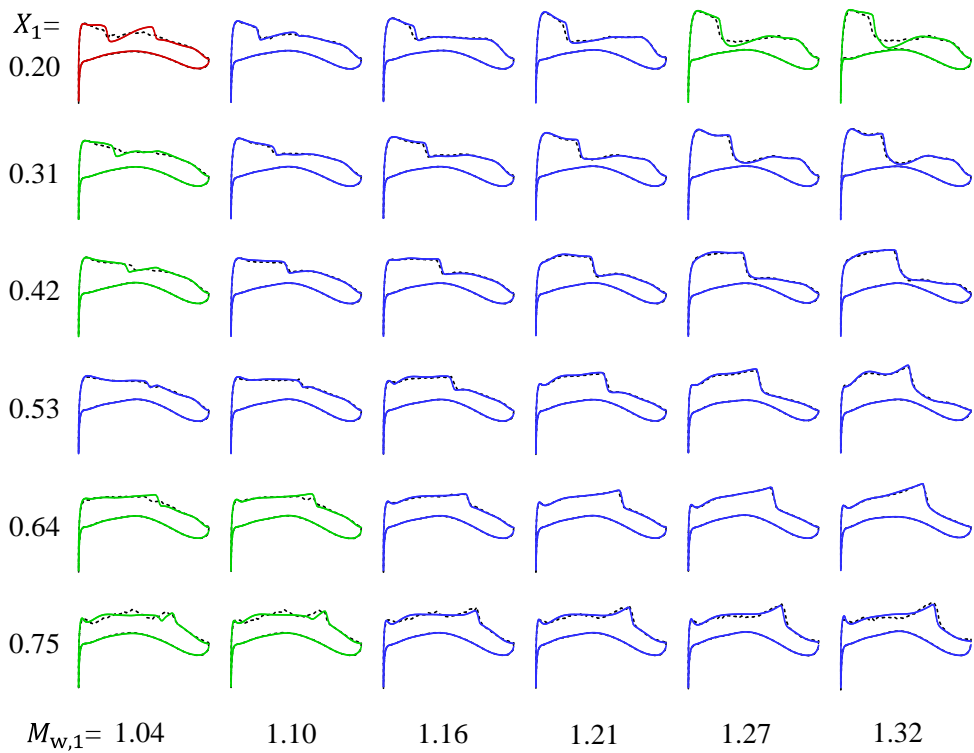

a) $M_w$ distribution matrix

**(Dashed: generated data; solid: CFD validation)**

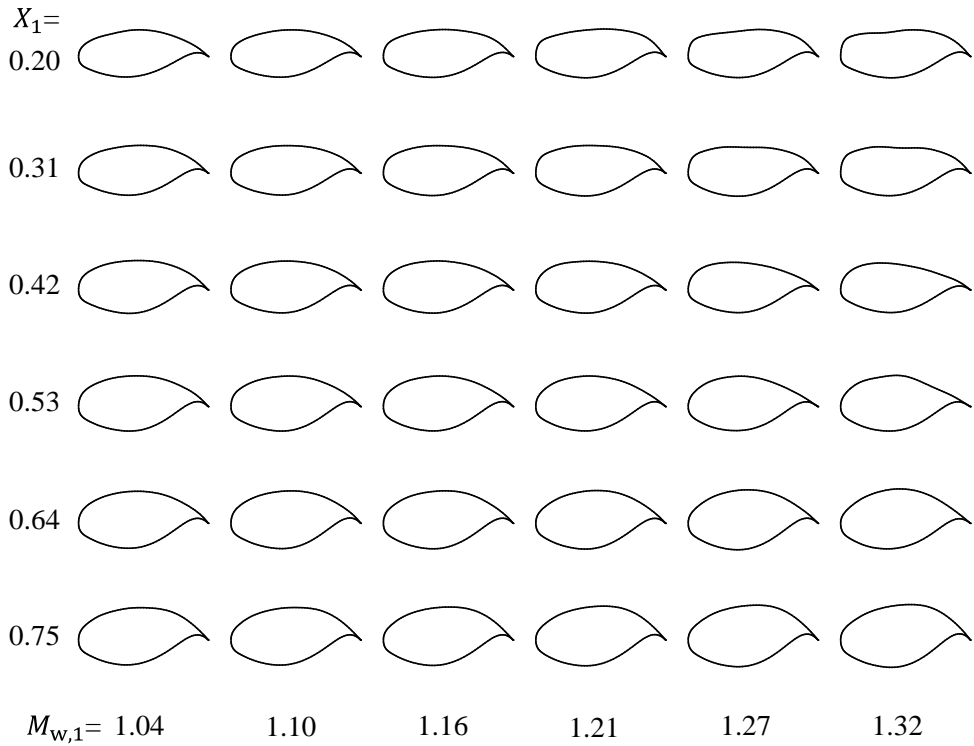

b) Geometry matrix

Fig. 18 Airfoil matrix of $M_{w,1} - X_1$ ($M_\infty$=0.73, $C_L$=0.75)

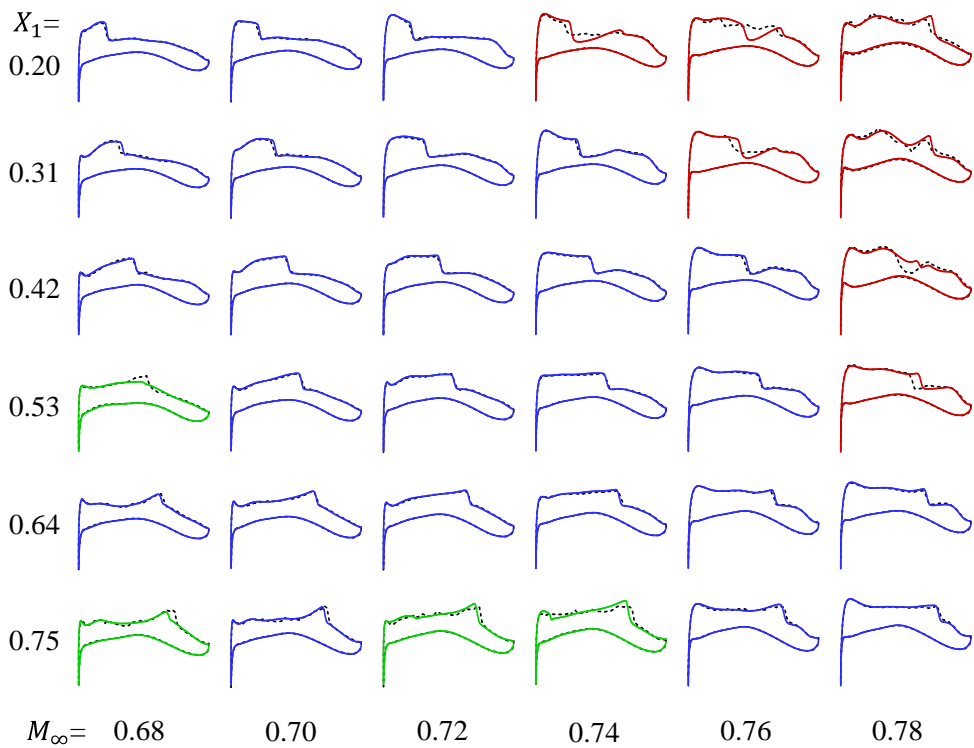

a) $M_w$ distribution matrix of $M_\infty - X_1$ ($C_L$=0.75, $M_{w,1}$=1.18)

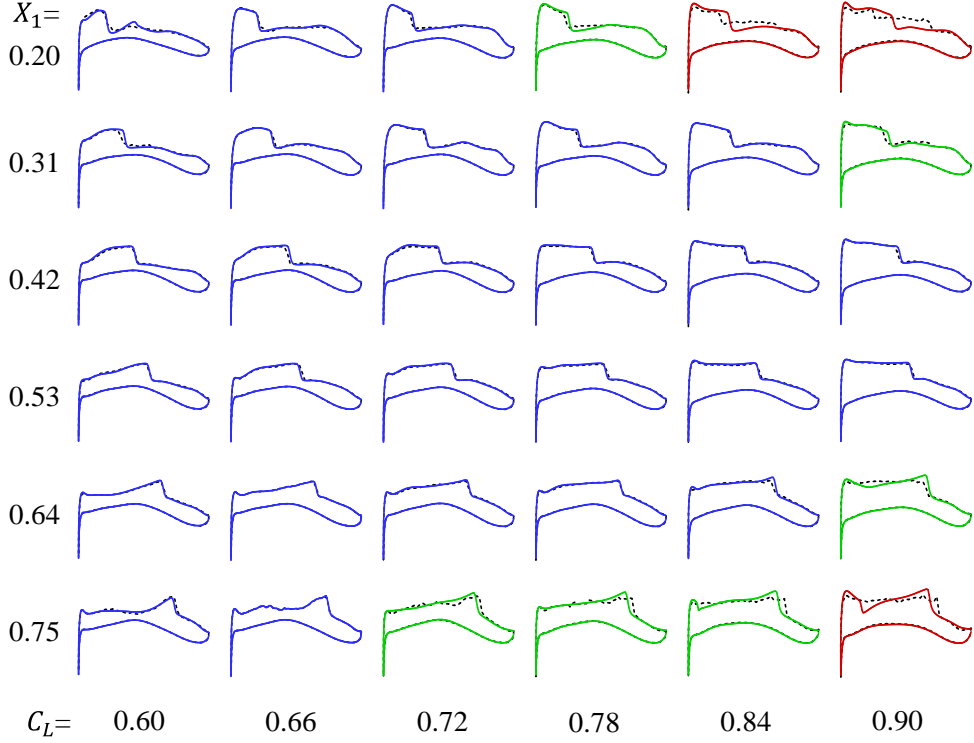

b) $M_w$ distribution matrix of $C_L - X_1$ ($M_\infty$=0.73, $M_{w,1}$=1.18)

**Fig. 19 Generated data (dashed) and CFD validation (solid)**

## V. Conclusion

Data representations greatly affect the performance and interpretability of machine learning models. Therefore, using physically interpretable features can benefit machine learning applications in fluid dynamics studies. This paper summarizes three requirements of physically interpretable features, namely, physical explainability, completeness, and disentanglement. A modification to variational autoencoders is proposed and theoretically proven to learn physically interpretable features. The proposed model, PIVAE, is also compared with other VAEs. Then, it is used for the inverse design of supercritical airfoils based on flow structure features instead of complete wall Mach number distributions.

The following conclusions have been reached.

(1) PIVAE is able to accurately reconstruct supercritical wall Mach number distributions with their latent variables, which are divided into physical codes and data features. The physical codes are equal to the specified flow structure features of the wall Mach number distribution, and the data features describe the rest of the information in the sample. Furthermore, PIVAE can generate new wall Mach number distributions based on given physical codes and random data features, and the actual flow structure features of the generated data are not affected by the data features.

(2) To train PIVAE, the sample set is required to provide the specified flow structure features as labels. The label distribution must be a uniform distribution so that the physical codes and data features of PIVAE are disentangled. In addition, cVAE is theoretically and numerically demonstrated to resemble PIVAE when using the same sample set.

(3) PIVAE can be used for inverse design of supercritical airfoils. The airfoil geometry can be directly generated along with the corresponding wall Mach number distribution, and the flow structure features of the airfoil can also be manipulated by changing the physical codes.

In summary, the proposed PIVAE can be used to extract physically interpretable features to improve the interpretability of machine learning models. It can also be used for inverse designs to manipulate flow structures.

## Acknowledgments

This work was supported by the National Natural Science Foundation of China under Grant Nos. 92052203, 91852108 and 11872230.